\documentclass{aa}  
\usepackage{float}
\usepackage{xcolor}
\usepackage{graphicx}
\usepackage{txfonts}
\begin{document}

   \title{Polarization aberrations in next-generation Giant Segmented Mirror Telescopes (GSMTs)}
    % Title brainstorming
    % II. Influence of segment-to-segment variations on polarization differential imaging
   \subtitle{II. Influence of segment-to-segment coating variations on high-contrast imaging and polarimetry}

   \author{Jaren N Ashcraft\inst{1,2,3}\fnmsep\thanks{NASA Hubble Fellow}
   \and Ramya M Anche \inst{1}
    \and Sebastiaan Y. Haffert \inst{1,4}
    \and Justin Hom \inst{1}
    \and Maxwell A. Millar-Blanchaer \inst{3}
    \and Ewan S. Douglas\inst{1}
     \and Frans Snik\inst{4}
    % \and Grant Williams\inst{1}
     \and Rob G. van Holstein\inst{5}
     % \and David Doelman\inst{6,4}
     \and Kyle Van Gorkom\inst{1}
    \and Warren Skidmore\inst{7}
    \and Manxuan Zhang\inst{3}}

   \institute{Steward Observatory, University of Arizona, 933N Cherry Avenue, Tucson, Arizona, 85721, USA \\ 
    \email{jashcraft@arizona.edu}
         \and James C. Wyant College of Optical Sciences, University of Arizona, 933N Cherry Avenue, Tucson, Arizona, 85721, USA        
         \and Department of Physics, University of California, Santa Barbara, CA, 93106, USA
          \and Leiden University, Niels Bohrweg 2, 2333 CA Leiden 
          \and European Southern Observatory (ESO), Alonso de Córdova 3107, Vitacura, Casilla 19001, Santiago de Chile, Chile
          \and TMT International Observatory LLC, 100 W. Walnut St., Suite 300, Pasadena, CA 91124, USA  }

   \date{Received October XX, 2024; accepted XXXX, XXXX}

% \abstract{}{}{}{}{} 
% 5 {} token are mandatory
 
  \abstract
   {Direct exo-earth imaging is a key science goal for astronomy in the next decade. This ambitious task imposes a target contrast of $\approx 10^{-7}$ at wavelengths from I to J-band. In our prior study, we determined that polarization aberrations can limit the achievable contrast to $10^{-5}$ to $10^{-6}$ in the infrared. However, these results assumed a perfect coronagraph coupled to a telescope with an ideal coating on each of the mirrors.}
   {In this study we seek to understand the influence of polarization aberrations from segment-to-segment coating variations on coronagraphy and polarimetry.}
   {We use the Poke open-source polarization ray tracing package to compute the Jones pupil of each GSMT with spatially-varying coatings applied to the segments. The influence of the resultant polarization aberrations is simulated by propagating the Jones pupil through physical optics models of coronagraphs using HCIPy.}
   {After applying wavefront control from an ideal adaptive optics system, we determine that the segment-to-segment variations applied limit the performance of coronagraphy to a raw contrast of approximately $10^{-8}$ in I-band, which is 2-3 orders of magnitude lower the target performance for high-contrast imaging systems on the ground. This is a negligible addition to the nominal polarization aberrations for ground-based systems. We further observe negligible degradation in polarimetric imaging of debris disks from segment-to-segment aberrations above and beyond the impact of nominal polarization aberration.}
   {}


   \maketitle
%
%-------------------------------------------------------------------

\section{Introduction}
\begin{figure*}[!ht]
\centering
\includegraphics[width=1\textwidth]{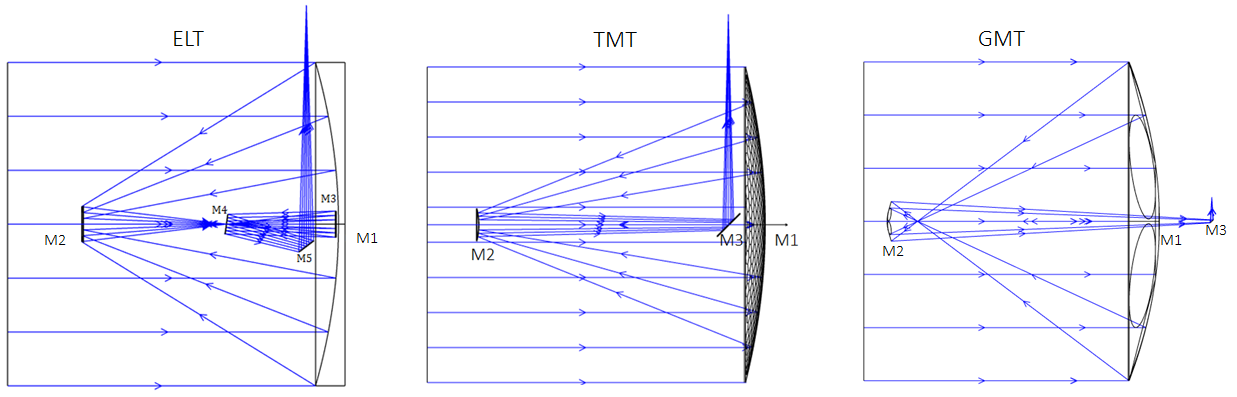}
\caption{Optical layout of the telescopes from Zemax\textsuperscript{\textregistered} for the three GSMT's. The TMT and ELT are cassegrain-type telescopes with a fold mirror configuration, whereas the GMT is Gregorian-type telescope with a fold mirror.}
\label{fig:opl-gsmt}
\end{figure*}

The next generation Giant Segmented Mirror Telescopes (GSMT's, shown in Figure \ref{fig:opl-gsmt}) will be the premier astronomical tools on the ground for direct imaging of faint exoplanets. The high angular resolution imaging enabled by their >25m apertures has the potential to expand the discovery space of exoplanets to include Earth-like planets. To fit in a reasonable volume, these large observatories employ fast primary mirrors ($\approx F/1$) and fold mirrors with high angles of incidence ($\approx 30^{\circ}-45^{\circ}$), which result in polarization aberrations that can limit coronagraphy.
\par
Polarization aberrations typically manifest as low-order aberrations (e.g., tilt, astigmatism) whose magnitude depends on the polarization state of the incoming light. Diattenuation is the aberration of the amplitude as a function of polarization, while retardance is the aberration of phase as a function of polarization \citep{chipman2018polarized,breckinridge2018terrestrial}. These can combine to result in aberrations of opposite sign for orthogonal polarization states. A formalism for polarization aberration was first introduced to astronomy in \cite{1992A&ASanchez_Martinez}. To understand the polarization aberrations manifesting as beam shifts in astronomical telescopes,  \cite{van2023polarization} describes the polarization aberrations for an inclined flat mirror in terms of Goos-Hänchen and Imbert-Federov beam shifts and discusses their implication on the coronagraphic performance. 

Although high-contrast imaging instruments employ extreme wavefront sensing and control, the adaptive optics (AO) system cannot simultaneously correct for aberrations in orthogonal polarization states \citep{breckinridge2015polarization}. Polarization aberrations are purely a function of the angle of incidence (AOI) and the coating on the optical surface. For a single layer, the diattenuation and retardance scale roughly quadratically with the AOI. However, most observatory mirrors are coated with a dielectric layer to enhance reflectivity or prevent oxidation of the primary reflecting layer, which can increase the retardance and degrade the coronagraphic performance.

The effect of polarization aberrations is far from new and has been observed in some of the current high-contrast imaging instruments in ground-based observatories. The SPHERE/ZIMPOL polarimetric imager has observed a shift between orthogonal polarization states that results in residual noise in the polarization differential imaging measurements as shown in \cite{Schmid_2018}. Subaru Telescope's SCExAO instrument and the SPHERE/IRIDIS instrument both requires substantive calibration of instrumental polarization to perform accurate polarimetry \citep{hart2021characterization,Zhang_2023, vanHolstein_2020}. The Gemini Planet Imager has shown that a differential astigmatism term between orthogonal polarization states is visible in polarized observations \citep{Blanchaer_2022}. These aberrations were incorporated during the design considerations of high-contrast imaging systems for next-generation space-based telescopes. For example, the design of the proposed HabEx \citep{gaudi2020habitable} and LUVOIR \citep{Will_polarization_luvoir} missions was driven partly to minimize the polarization aberrations that degrade the coronagraphic performance. \cite{balasubramanian2005polarization} developed optimized mirror coatings to reduce the polarization aberrations for the proposed Terrestrial Planet Finder Coronagraph (TPF-C). Further, \cite{luo2020effects} simulated the polarization aberrations for a 2 m, F/14 three-mirror anastigmat (TMA) telescope to estimate the PSF ellipticity affecting the measurements of weak gravitational lensing. \cite{anche2023simulation,doelman2023falco} have performed simulations of high-contrast polarimetric observations of disks and planets for the upcoming Nancy Grace Roman Coronagraph instrument \citep{kasdin2020nancy} to show that polarization aberration residuals are below the expected coronagraphic performance.  

For the proposed future Habitable Worlds Observatory (HWO) aiming to achieve a raw contrast of $10^{-10}$, minimizing the polarization aberrations is expected to be one of the driving requirements. As a preliminary analysis for the HWO, \cite{anche2023estimation} modeled polarization aberrations for 6m on-axis and off-axis TMA designs using different coating recipes for the mirrors and demonstrated that the coating recipe (refractive index and thickness of the protective coating) plays a vital role in the polarization aberrated coronagraphic performance. Thus, modeling polarization aberrations and estimating their effect on the contrast is critical for next-generation telescopes with high-contrast imaging instruments.

The first paper in this series on polarization aberrations in GSMTs, \cite{GSMTsI_2023}, demonstrated the effect of polarization aberrations on coronagraphic performance by presenting simulations of a static polarization model of the telescope with perfect coronagraphs (PC). We refer the reader to \cite{GSMTsI_2023} for a complete description of the simulations. Below, we summarize only the model parameters and important findings of the study.

\section{Review of previous findings} 
We model and perform the raytracing of the GSMTs using Zemax\textsuperscript{\textregistered} and obtain Jones pupils using the Python-based polarization ray tracing package (PRT), Poke \citep{Ashcraft_poke_2022, Ashcraft_poke_2023}. The TMT and ELT mirror coating were a 4-layer Gemini-like coating \citep{Schneider2016}, and bare Aluminum was used on GMT. We simulated the Jones Pupils at the exit pupil for each of the telescopes using PRT for astronomical filter bands \textit{U} to \textit{N}. The eigenvalues of the Jones pupils were used to estimate diattenuation and retardance. Among the three GSMTs, GMT showed the highest diattenuation, followed by TMT and ELT, while TMT showed the highest retardance, followed by ELT and GMT. The retardance of all three telescopes reduces with increasing wavelength. On the other hand, diattenuation for GMT showed an increasing trend in optical and near-IR wavelength regimes due to the refractive index variation of bare aluminum. The high retardance for TMT and ELT is attributed to the 4-layer Gemini-like coating. 
\par The wavefront sensing and control system can compensate for the common aberrations between the \textit{X} and \textit{Y} polarized light. Thus, they are removed from the Jones pupils $\phi xx$ and $\phi yy$ terms (see Figure 5 in \citealt{GSMTsI_2023}) and propagated through the perfect coronagraphs that remove the electric field modes based on their spatial order \citep[2, 4, and 6;][]{cavarroc2006coronagraph, guyon2006coronagraph}. The achievable peak raw contrast for all three telescopes is limited by the dominant polarization aberrations: retardance tilt (i.e. beam shifts), and retardance defocus (i.e. differential astigmatism). For a brief introduction to these aberrations, please refer to Section 1 of \cite{GSMTsI_2023}. As these polarization aberrations depend on the refractive index variation, the chromatic behavior is seen in the residual images for each telescope over the different filter bands. 
\par The peak raw contrast is calculated using the coronagraphic residual images for each of the telescopes at each filter band. The high-contrast imaging instruments onboard the GSMTs are expected to obtain a peak contrast of $10^{-5}$ - $10^{-6}$ in the R and I bands to search for biomarkers such as oxygen A-band at 720nm \citep{kasper2021,males2022conceptual}. However, simulations in \cite{GSMTsI_2023} show that the peak raw contrast is $>$ $10^{-5}$ for all the three GSMTs in the \textit{U}- \textit{y} bands for the 2nd and 4th order coronagraphs, which is an order of magnitude worse than the required performance. The peak raw contrast in the order of $10^{-5}$ - $10^{-6}$ is obtained only for the \textit{J} - \textit{N} bands. 
\par Among all three GSMTs, the ELT shows the worst contrast degradation due to polarization aberrations because it has five mirrors coated with a 4-layer coating and the fastest primary (F/0.7) compared to TMT (F/1). Although GMT also has a fast primary (F/0.87), it outperforms the TMT and ELT by an order of magnitude as its mirrors are coated with bare aluminum and therefore have the lowest retardance. Models in which the GMT mirrors are coated with bare silver and Gemini-like coating yield a peak raw contrast performance of GMT that is very similar to that of ELT and TMT.
Thus, coating plays a significant role in the magnitude of the polarization aberrations, and optimized coating recipes would reduce the aberrations \citep{van_Holstein_2023}. The previous modeling and simulations were performed using an ideal coating recipe on all the telescope mirrors, including primary mirror segments for all three GSMTs. Additionally, for GMT mirrors coated with bare aluminum, the influence of $Al_2O_3$ on bare aluminum was not considered, which has been observed before in aluminum-coated telescope mirrors \citep{van_Harten_2009,sankarasubramanian1999measurement}.  
\par In this study, we aim to bring these findings to the next level of fidelity by examining how segment-to-segment coating variations will influence high-contrast imaging and polarimetry. In Section \ref{sec:segment_var}, we develop the modeling of the GSMTs subject to segment-to-segment variations using integrated PRT and diffraction models. In Section \ref{sec:coronagraphy}, we use these models to re-evaluate the coronagraphic performance of the GSMTs subject to these effects by statistically probing the possible space of segment-to-segment variations. In Section \ref{sec:polarimetry}, we use these models to understand how polarization aberrations influence polarimetric imaging of distant point sources and debris disks. We summarize our findings and provide recommendations in Section \ref{sec:Summary}.

%-------------------------------------------------------------------

%-------------------------------------------------------------------

\section{Modeling approach}
\label{sec:segment_var}
\begin{figure*}[!ht]
    \centering
    \includegraphics[width=1\textwidth]{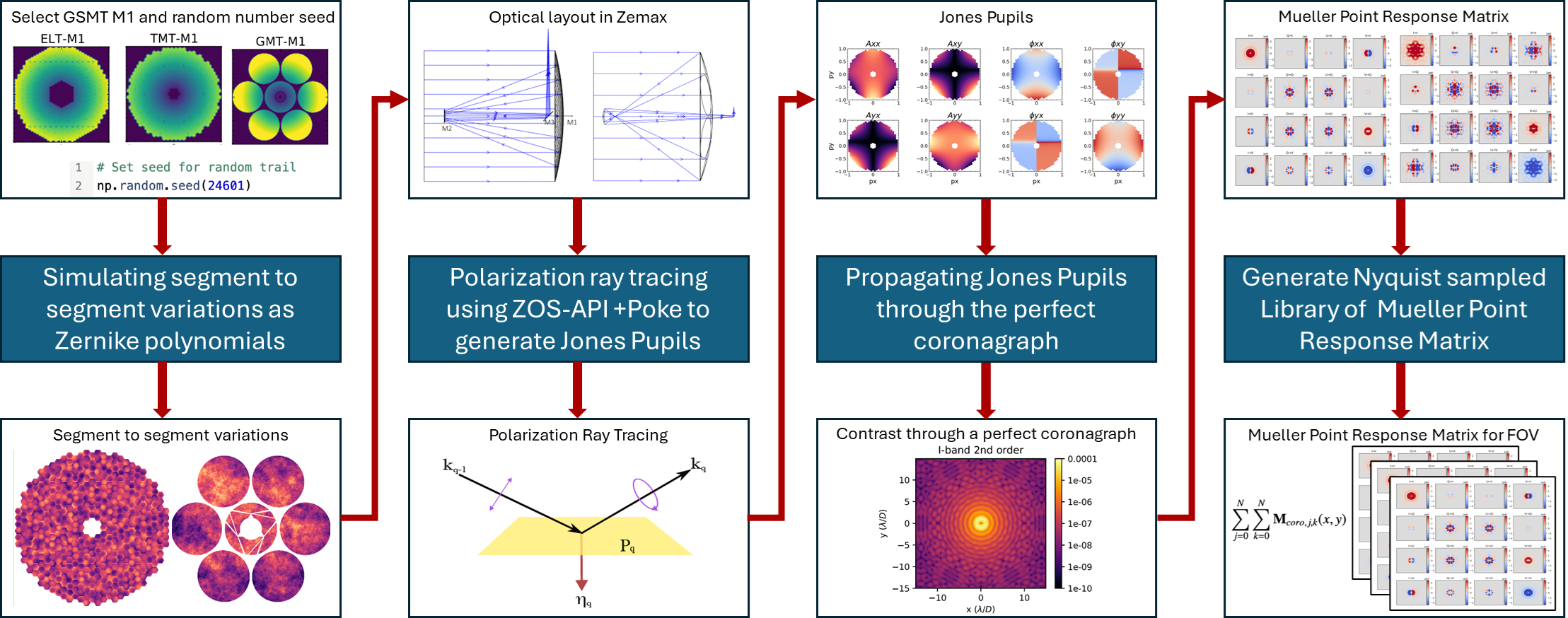}
    \caption{Steps to simulate polarimetric imaging of a generally off-axis source in the presence of polarization aberrations through the GSMTs is described here. As a first step, we simulate the segment-to-segment coating variation to be a sum of the piston, tilt, and focus terms of the Zernike polynomials. We then perform the PRT using Poke to estimate Jones pupils for all three telescopes. The Jones pupils and the corresponding low-order polarization aberration terms that are fit to these Jones pupils are available in \cite{anche2023estimation}. The Jones pupils are propagated through the perfect coronagraphs to calculate the residuals and the peak raw contrast. The Jones pupils are further used to estimate the Amplitude Response Matrix and Mueller Point Response Matrix for use in extended scene simulation.}
    \label{fig:model-sch}
\end{figure*}

In this work we simulate the influence of coating variations on polarization aberrations with PRT. PRT is a method of propagating the complex amplitudes of orthogonal polarization states through three dimensional optical systems. We review this technique in Section 3 of our previous work \citep{GSMTsI_2023}. This review is not exhaustive for readers that may be interested in reproducing the PRT algorithm, so we also refer readers to Chapters 10 and 11 of \cite{chipman2018polarized}. To perform PRT we use the Python package Poke: an open-source ray-based physical optics platform (described in \citealt{Ashcraft_poke_2023}). In our previous study we built PRT into Poke to compute the Jones pupil of the GSMTs. Poke directly interfaces with other open-source physical optics packages like High Contrast Imaging in Python \citep[HCIPy,][]{por2018hcipy}, which we use to simulate the diffraction response of a coronagraph to polarization aberrations. Poke can perform PRT with arbitrary thin-film multilayer stacks with spatial variation, which will be key in simulating spatially-varying coating thicknesses on the GSMTs. The code used to conduct the simulations in this study can be found at \cite{gsmts_sim}, and we illustrate its operation in Figure \ref{fig:model-sch}.

Understanding what shape the coating thickness variation will take is somewhat difficult because of the challenge of measuring coating thicknesses across meter-scale segments. However, we can begin to understand the spatial distribution of a given coating by looking to work conducted by Gemini Observatory. The Gemini primary mirror coating chamber uses magnetron sputtering to deposit their protected silver coating. Magnetron sputtering operates by applying an electrical bias on the material to deposit (called the "target," e.g. Aluminum, Silver). The bias attracts the plasma, resulting in high-energy collisions that free matter from the target and allow it to fly toward the substrate being coated. The deposition rate of this coating process is high, but difficult to conduct with spatial uniformity on large optics. Investigators at Gemini telescope studied witness samples from their coating chamber ellipsometrically to determine the coating thickness deposited across a substrate, and found a $15\%$ (or $1.3$nm) peak-to-valley variation in the deposited coating over a 4cm $\times$ 4cm substrate \citep{Schneider_Gemini_Coating}. When considering the geometry of typical coating chambers, we know that the predominant modes of nonuniform coating deposition appear as a mixture of tilt and focus-like shapes \citep{bishop_xray_sputter_2019}. Furthermore, the TMT is considering a similar coating recipe and procedure, and expects variations in the coating thickness as large as $10-20\%$ \citep{private_communication_tmt}.

\begin{figure}[h]
    \centering
    \includegraphics[width=0.5\textwidth]{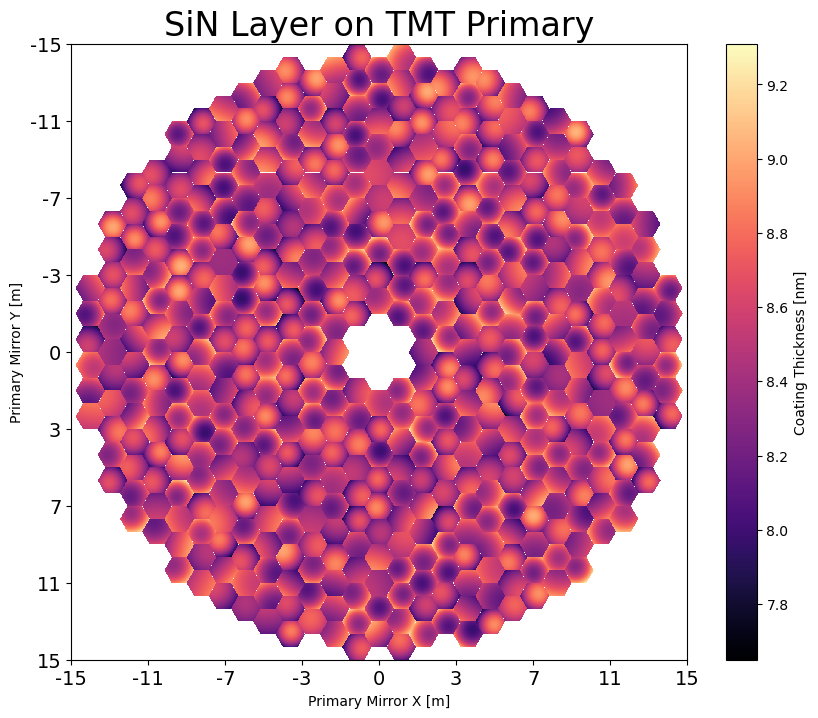}
    \caption{Example of low-order coating variation on the TMT primary mirror. We generate a random set of Zernike coefficients from Z1-Z4 and apply it to each segment using the prysm numerical optics package. The error is normalized to have a peak-to-valley of $10\%$ of the nominal thickness and then added to the nominal thickness. Quantifying the impact of these errors on coronagraphy and polarimetry is important to understanding the as-built performance of the GSMTs.}
    \label{fig:coating_variation_tmt}
\end{figure}

\begin{figure}[h]
    \centering
    \includegraphics[width=0.5\textwidth]{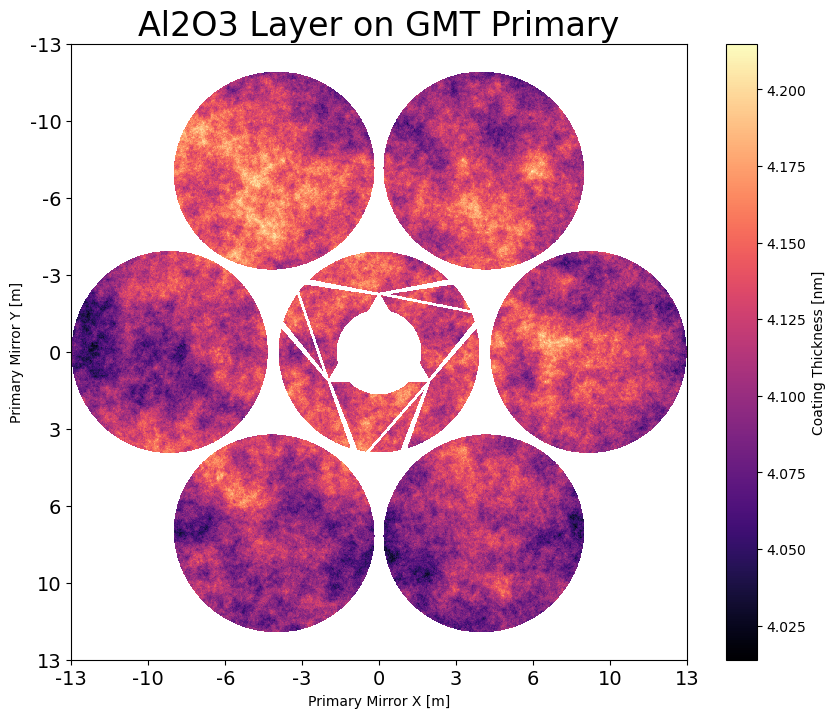}
    \caption{Example of the PSD-generated coating variation assigned on the GMT primary mirror. We generate a power law with a given index $n$ and then enforce its peak-to-valley variation to be consistent with the $\pm 0.08nm$ from \cite{van_Harten_2009}. Varying the PSD index $n$ allows us to parameterize the influence of the Aluminum Oxide layer on the GMT.}
    \label{fig:coating_variation_gmt}
\end{figure}

To model these effects, we consider the dielectric coating on each mirror segment to be a sum of the piston, tilt, and focus terms of the Zernike polynomials. These thickness variations are applied to the overcoat layer of the ELT and TMT (illustrated in Figure \ref{fig:coating_variation_tmt}). The coating thickness variation on the underlying reflective layer for the GSMTs will predominantly manifest as a wavefront aberration that is common to all polarization states, which will be removed by the AO system. Therefore, we assume an ideally deposited reflective layer. The GMT will not have a dielectric overcoat layer, but \cite{van_Harten_2009} showed that the aluminum oxide layer that grows on bare aluminum substrates could result in negative consequences to polarimetry. The spatial distribution of this growth is unknown, so we elect to simulate it with a power law whose power spectral density ($PSD$) is defined by the power law in Equation \ref{eq:power_law}\footnote{This is the default PSD for the SurfaceAberration class in HCIPy},

\begin{equation}
    PSD = r^{n},
    \label{eq:power_law}
\end{equation}

where $r$ is the radial spatial frequency and $n$ is the power law index. \cite{van_Harten_2009} suggests that the $Al_{2}O_{3}$ layer approaches a thickness of $4.12 \pm 0.08 nm$ roughly 800 hours after being coated, but the result is computed from an ellipsometer which spatially averages the coating. Therefore we will assign the power law distribution with a peak-to-valley equal to the uncertainty in the measurement. One such resulting surface from this approach is shown in Figure \ref{fig:coating_variation_gmt}. The power law index $n$ is also not well-understood for telescope primary mirrors, so we simulate several cases in this study to parameterize the influence of this quantity. Given the uncertainty in the dielectric layer for the TMT and ELT, we also consider two cases where the film peak-to-valley film variation is $20\%$ and $50\%$. Table \ref{tab:coating_var} summarizes the thickness variation we simulate for each of the GSMTs. Note that we hold the GMT peak-to-valley variation the same, but instead vary the power law index $n$.

\begin{table}[H]
    \centering
    \begin{tabular}{c|c c c|c}
    \hline
    \hline
        &  & Thickness Variation &  & \\
        GSMT & Case 1 & Case 2 & Case 3 & Material  \\
        \hline 
        GMT & $n = -1$ & $n = -2$ & $n= -3$ & $Al_{2}O_{3}$ \\
        ELT & $10\%$ & $20\%$ & $50\%$ & $Si_{3}N_{4}$ \\
        TMT & $10\%$ & $20\%$ & $50\%$& $Si_{3}N_{4}$ \\
        \hline
    \end{tabular}
    \caption{Parameters for spatial variations used for each case studied for each GSMT. The peak-to-valley aberration applied is symmetric about zero and, as such, has positive and negative values. These randomly generated screens are then added to the nominal thicknesses described in \cite{GSMTsI_2023}.}
    \label{tab:coating_var}
\end{table}

%--------------------------------------------------------------------
% Hypothesis, the segment variations are higher-order erros that the order 6 PC can't fully reject
\section{Effect on coronagraphy}
\label{sec:coronagraphy}

Since the coronagraph architectures for the GSMTs are not finalized, we repeat the procedure from our previous study and employ the Perfect Coronagraph (PC) models described in \cite{guyon2006coronagraph} and \cite{cavarroc2006coronagraph}. These are readily accessible with polarized field propagation in HCIPy and described in Section 6 of \cite{GSMTsI_2023}.

\begin{figure}[H]
    % \centering
    \includegraphics[width=0.45\textwidth]{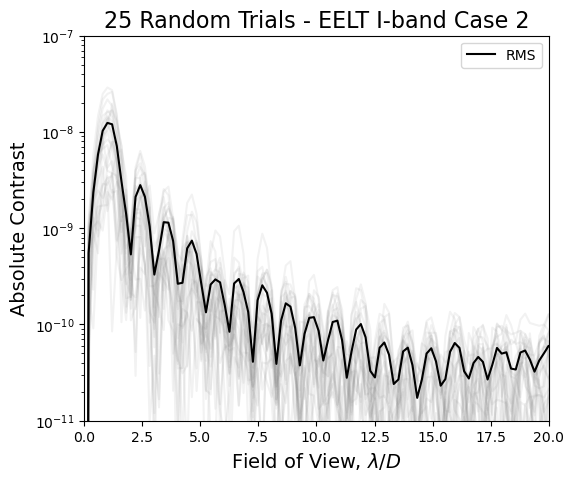}
    \caption{Radial profile of the azimuthally averaged focal plane residuals given the polarization aberrations of spatially-varying coatings on the ELT primary mirror. These data are generated by taking the difference of coronagraphic images with and without segment-to-segment coating variations. These data are computed in I-band for a PC of order 2, and plotted on a log scale to show the the change in the absolute value of contrast and highlight the distribution's features. Each individual trial is plotted in gray, and the RMS of these data is plotted in solid black.}
    \label{fig:rand_profiles}
\end{figure}

We statistically sample the random realizations of the thickness variations specified by Table \ref{tab:coating_var} using a set of 25 randomized trials for each Case, GSMT, and PC order (2, 4, 6). An example of one such set of trials is shown in Figure \ref{fig:rand_profiles}, where we simulate Case 2 for the ELT equipped with a 2nd-order PC. We then subtract off the coronagraphic image assuming a perfectly uniform coating to arrive at the change in contrast introduced by the spatially-varying coating. The change in contrast is observed to be small (maximizes near $1 \lambda / D$ at $\approx 10^{-8}$), and the plot of the RMS roughly envelopes the individual trials. For another example in 2D, we show the log-scaled difference of one realization of these PSFs for the GMT with the nominal polarization aberrated PSF subtracted in Figure \ref{fig:alumina_influence}. This Figure shows the influence of increasing power law index $n$, as well as the structure of the residual aberration.

\begin{figure*}
    \centering
    \includegraphics[width=\textwidth]{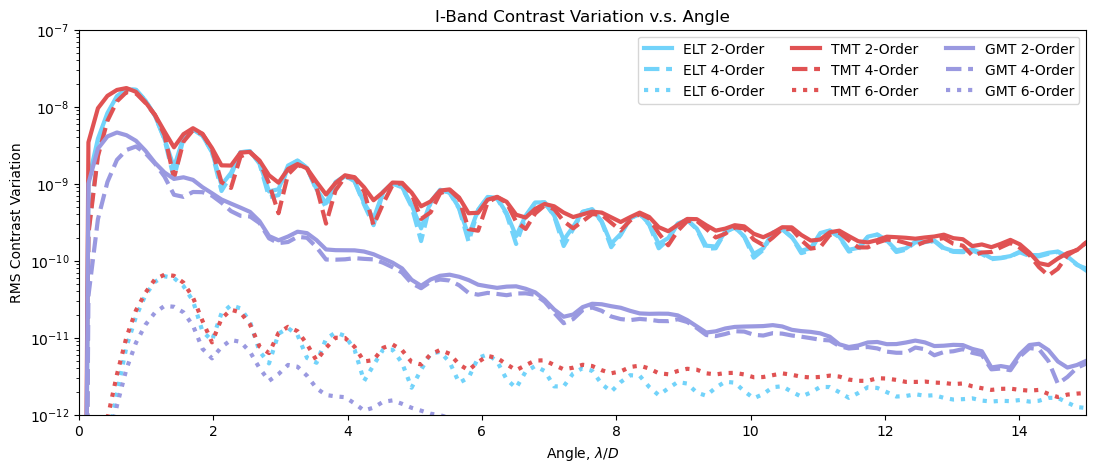}
    \caption{Plot of the RMS contrast variation as a function of angular separation for Case 2 in I-band. The lines represent the azimuthal average of the standard deviation of the simulated coronagraphic residuals. One can interpret these data as the anticipated contrast degradation due to coating variations. We observe that order 2-4 coronagraphs do not have substantially different performance, with an RMS contrast variation near the inner working angle of $\approx 10^{-8}$ for the TMT and ELT, and $\approx 5 \times 10^{-9}$ for the GMT. Order 6 coronagraphs have a much tighter standard deviation, which is below $10^{-10}$ contrast. These data suggest that segment-to-segment errors will not be a dominant effect that limits high-contrast imaging. For the curves that represent Case 1 and 3, please refer to the appendices.}
    \label{fig:psf_residuals}
\end{figure*}

\begin{figure*}
    \centering
    \includegraphics[width=1\textwidth]{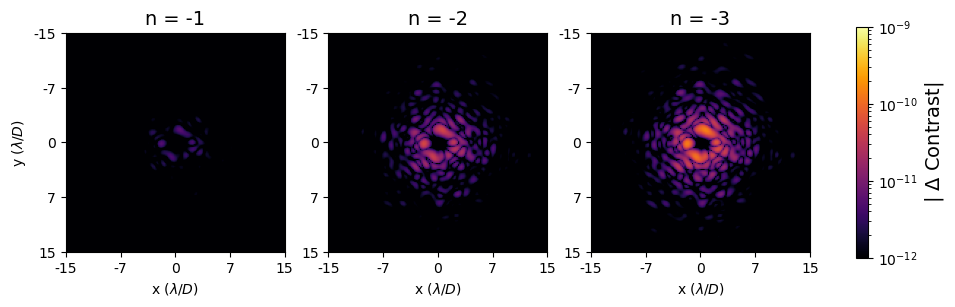}
    \caption{The influence of the $Al_{2}O_{3}$ layer on coronagraphic performance of the GMT for a 6th-order PC in I-band. These data show the difference between the case \emph{with}  and without segment variations.  Here $n$ is the exponent of the PSD used to generate the distribution of $Al_{2}O_{3}$ on the primary mirror. We observe that the coronagraphic residuals generally increase with decreasing $n$, but do not exceed $10^{-9}$ contrast.}
    \label{fig:alumina_influence}
\end{figure*}

In Figure \ref{fig:psf_residuals} we show the RMS of the Case 2 trials for all GSMTs and coronagraph orders. These data represent the expected variability in contrast that is created by applying spatially non-uniform errors across each mirror segment, and correcting the resulting scalar wavefront error with the AO system. We observe that order 2 and 4 coronagraphs behave very similarly and that the order 6 coronagraph performs orders of magnitude better, which is consistent with our results in \cite{GSMTsI_2023}. For Case 2, the expected variability peaks with the TMT at $1.7 \times 10^{-8}$ contrast. To put this in context, the signal from the segment-to-segment errors is 100 times fainter than the AO residuals projected for high-contrast imaging instruments on these systems. This is a very encouraging result, because it suggests that nonuniform coatings on the primary mirrors of the GSMTs will not be a limiting factor in high-contrast imaging. The same Figures for Case 1 and 3 can be found in Appendix \ref{sec:appendix_a}, but the worst case scenario shows that the peak residuals behind an order 2 coronagraph are $4.3 \times 10^{-8}$, which will nominally not be a substantive error for high-contrast imaging. We therefore conclude that while polarization aberrations will overall be an effect that dominates contrast in the near IR and shorter wavelengths, the segment-to-segment variation will not create an additional dominant source of error. However, given that these RMS contrast levels are orders of magnitude larger than the target contrast for HWO of $\approx 10^{-10}$, we recommend that a similar analysis be conducted to assess HWO's sensitivity to segment-to-segment variations.

%It is worth noting that this variation may contribute to the theoretical contrast floor achievable by polarization differential imaging instruments
% Between the three cases studied, the peak contrast degradation for the TMT and ELT order 2 coronagraphs doesn't change by greater than an order of magnitude. The GMT on the other hand changes by greater than an order of magnitude

%--------------------------------------------------------------------

\section{Effect on polarimetry}
\label{sec:polarimetry}
While the segment-to-segment variations will not be the limiting factor in high-contrast imaging for the GSMTs, another area critical to study is how polarization aberrations from coating variations impact polarimetric observations. 

To understand the response of a telescope's point-spread function to polarization aberrations, we construct the Mueller Point-Response Matrix (MPRM). Using PRT we compute the Jones Pupil $\mathbf{J}(x,y;\theta)$ in the local coordinate system of the coronagraph, which is computed for some field angle $\theta$. We then propagate $\mathbf{J}(x,y;\theta)$ through the coronagraph ($\mathbf{C}$) to arrive at the amplitude response matrix $\mathbf{A}_{coro}(x',y';\theta)$, as shown in Equation \ref{eq:coro_prop},

\begin{equation}
    \mathbf{A}_{coro}(x',y';\theta) = \mathbf{C}[\mathbf{J}(x,y;\theta)e^{-i\frac{\phi_{xx} + \phi_{yy}}{2}}].
    \label{eq:coro_prop}
\end{equation}

Here $(x',y')$ represents the coordinate system of the focal plane, and the exponential term is the phase correction applied by the ideal AO system. The $\mathbf{A}_{coro}(x',y';\theta)$ is the response of the coronagraphic field, but to understand its influence on the detected intensity, we must transform it to a Mueller matrix. This operation is shown in Equation \ref{eq:jones_to_mueller},

\begin{equation}
    \mathbf{M}_{coro} = \mathbf{U} (\mathbf{A}_{coro} \otimes \mathbf{A}_{coro}) \mathbf{U}^{-1}.
    \label{eq:jones_to_mueller}
\end{equation}

Where $\otimes$ denotes the Kronecker product and $\mathbf{U}$ is given by Equation \ref{eq:jtm_U},

\begin{equation}
    \mathbf{U} = 
    \begin{pmatrix}
        1 & 0 & 0 & 1 \\
        1 & 0 & 0 & -1 \\
        0 & 1 & 1 & 0 \\
        0 & i & -i & 0 \\
    \end{pmatrix}.
    \label{eq:jtm_U}
\end{equation}

\begin{figure*}
    \centering
    \includegraphics[width=0.75\textwidth]{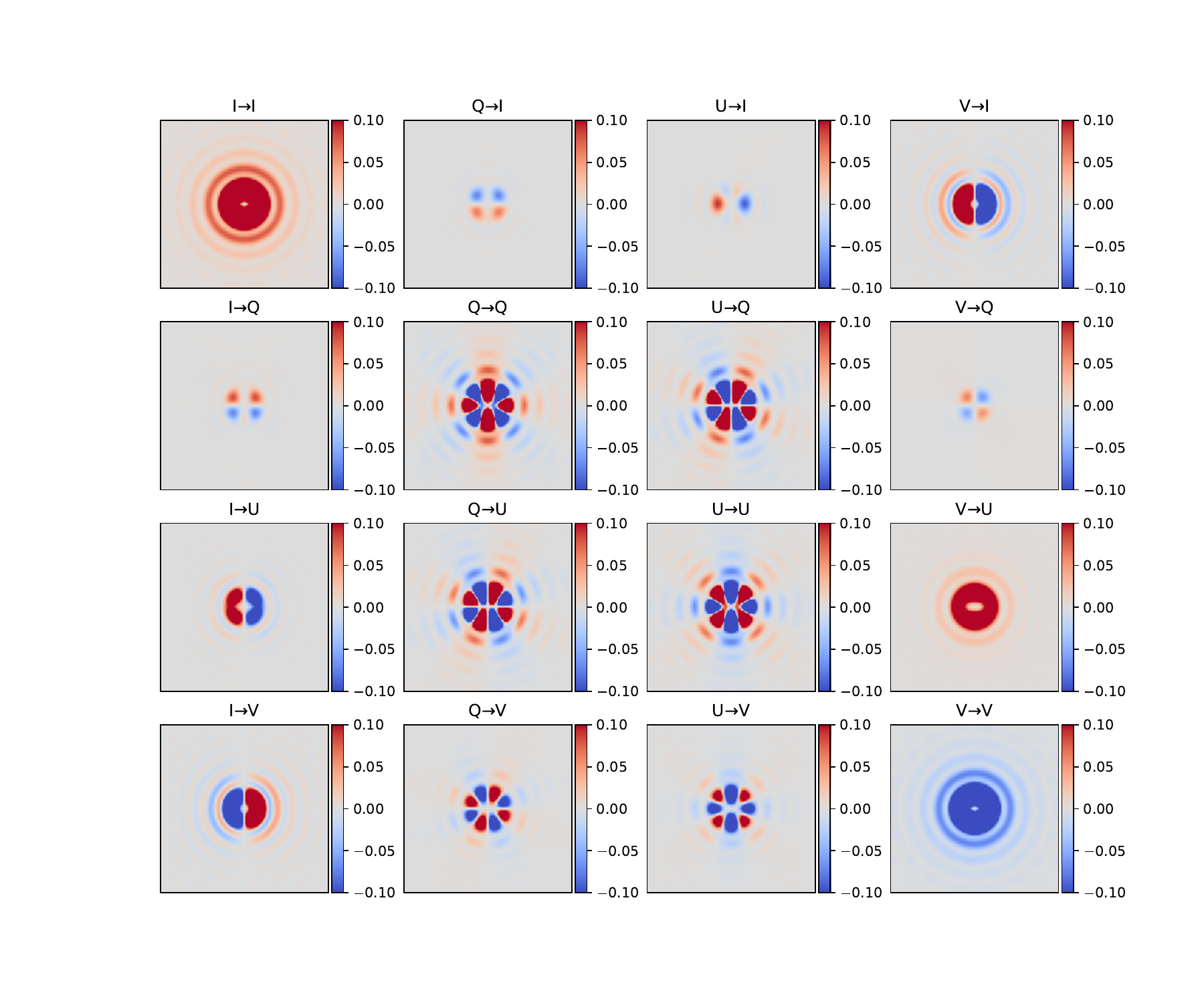}
    \caption{Figure showing the Mueller PRM $\mathbf{M_{coro}}$ for the TMT in I-band to show how polarization shapes the PSF. A simple imager would observe the $I\rightarrow I$ component of $\mathbf{M}_{coro}$. The data shown here are normalized to the peak of this component. In response to unpolarized light, the observed image of a point-source is simply the sum of the top row times the Stokes vector that represents the source. However, for images that we detect in linear polarized light the $U\rightarrow Q$ and $Q \rightarrow U$ retardance terms may sculpt the disk image in a way that inhibits precise detection. The data here is plotted with a colorbar such that bright signals are saturated to enhance the structure in the first lobe of the PRM elements.}
    \label{fig:tmt_prm}
\end{figure*}

\begin{figure*}
    \centering
    \includegraphics[width=0.75\textwidth]{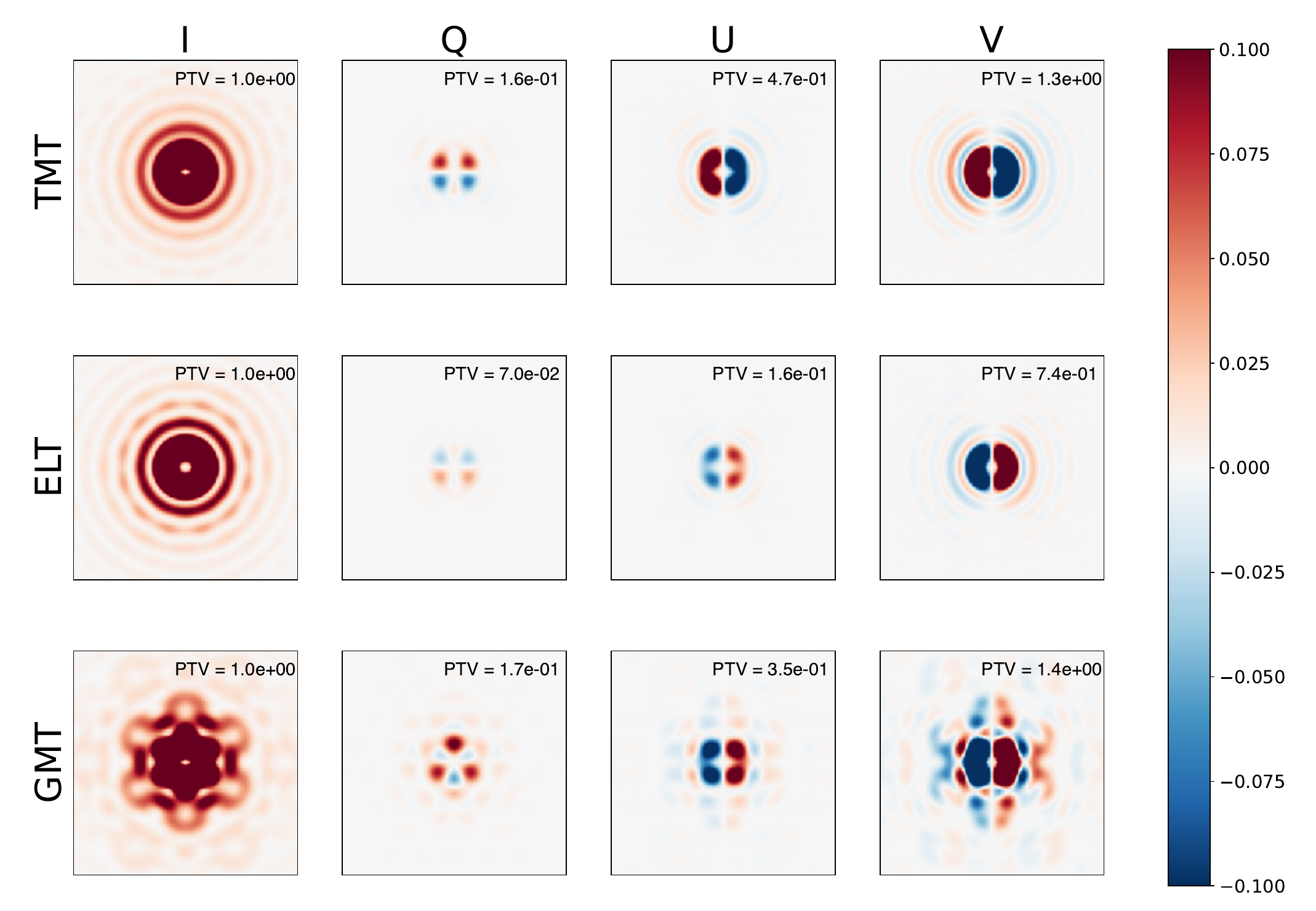}
    \caption{The Stokes vectors that would be detected by a perfect high-contrast polarimeter (order = 2 PC) that observed an unpolarized star subject to the polarization aberrations of the GSMTs. These data are normalized to the peak of the $I\rightarrow I$ component. Each row is a Stokes vector that belongs to one of the GSMTs; (Top) TMT, (Middle) ELT, (Bottom) GMT. These data are plotted on a scale to emphasize the contribution of the polarization aberration to the stellar leakage at the inner working angle. The peak-to-valley (PTV) of the Stokes images is shown on the top-right of each image. Some terms have a PTV larger than 1 because they can take on negative values.}
    \label{fig:stokes_unpolarized_star}
\end{figure*}

$\mathbf{M}_{coro}$ describes the response of the focal plane intensity to some incoming Stokes vector $S = [I,Q,U,V]^{T}$ subject to the polarization aberrations present in our instrument. An example of $\mathbf{M_{coro}}$ is shown in Figure \ref{fig:tmt_prm}, where we show the MPRM for the TMT in I-band. In response to an unpolarized star, i.e. $S = [1,0,0,0]^{T}$, a detector would measure the top left element of this matrix ($I\rightarrow I$). However, in the presence of polarization aberrations some of the unpolarized power $I$ becomes linearly ($Q, U$) and circularly ($V$) polarized, which can be detected by polarimeters. This is described by the polarizance vector of $\mathbf{M}_{coro}$, which is the leftmost column of the matrix. Shown in Figure \ref{fig:stokes_unpolarized_star} is what a theoretical, uncalibrated, full-Stokes polarimeter would measure from each of the GSMTs in response to an unpolarized star. The polarization aberrations manifest as structure near the inner working angle whose shape and orientation is a function of the incident polarization state. When observing targets at small angular separations to a host star using polarimetric differential imaging, this could be a limiting factor in the best achievable polarimetric contrast. 

To best understand the influence of polarization aberrations on the ability to perform high-contrast polarimetry, we consider other extended sources that can be detected in polarized light using polarization differential imaging (PDI). This includes circumstellar disks and reflected light from exoplanets, which may be limited by the structure imparted by the MPRM. In contemporary systems, the influence of first-order polarization aberrations (instrumental polarization, cross-talk) on polarimetry can in part be accounted for using calibration routines.

Calibration routines for current high-contrast polarimeters involve a combination of laboratory characterization, internal source calibration, and on-sky standard observations. Observations of unpolarized and polarized standards can retrieve the retardance and diattenuation of telescope mirrors, while internal calibration sources can determine the polarimetric properties of components downstream of the telescope. All polarizing components are represented as 4 $\times$ 4 Mueller matrices, and the overall instrument is represented as the product of such components. To characterize instrumental polarization from the telescope, observations of standard stars at several altitudes on the same night as science observations is a robust method of estimating the mirrors' physical properties. As is the case with SPHERE-ZIMPOL \citep{Schmid_2018}, SPHERE-IRDIS \citep{vanholstein2020sphere, deBoer_sphere_2020}, and SCExAO-VAMPIRES \citep{Zhang_2023}, internal calibrations are preferably also performed at a regular cadence (e.g. weekly, monthly) to track any changes in polarimetric properties. Using internal-source and on-sky calibration datasets, polarimeters such as SPHERE-IRDIS have achieved absolute polarimetric error on the order of ${0.1\%}$ \citep{vanholstein2020sphere} and a polarized contrast of $10^{-5}-10^{-6}$ \citep{VanHolstein_polarizedcontrast_2017}. In \cite{van_Holstein_2021}, SPHERE-IRIDIS achieved a $1-\sigma$ polarized contrast of $10^{-8}$ at 1.5 arcsec. In this section we construct a model of a debris disk and subject it to the polarization aberrations that arise in segments with coating variations to assess the degree of polarimetric accuracy that needs to be calibrated, and assess the presence of any spatially-varying structure that might arise.

%This will likely change if we choose to do a smaller disk. I will also pick more "uniform" parameters since we no longer care about making a TWA 7 model

\begin{figure*}
    \centering
    \includegraphics[width=\textwidth]{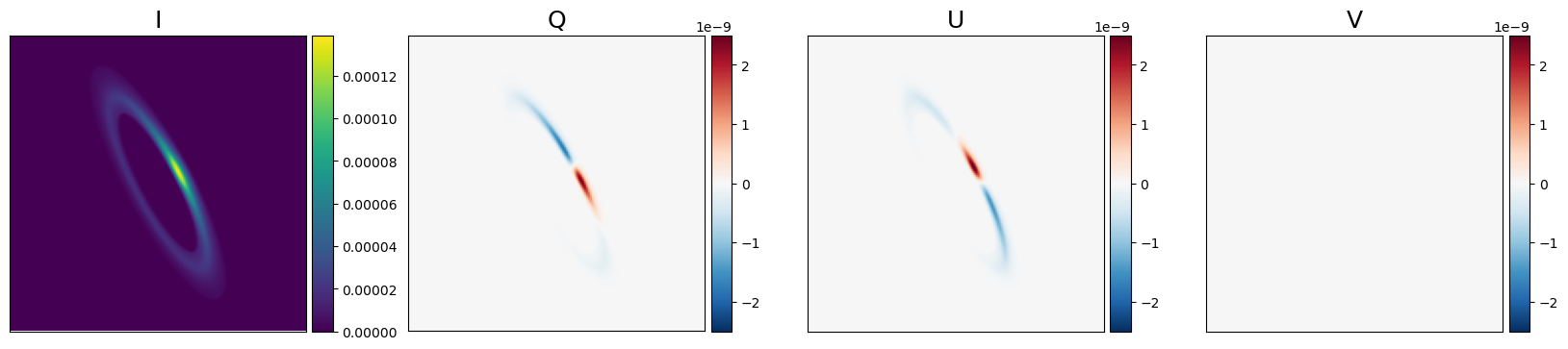}
    \caption{The nominal Stokes image of the debris disk model. I is the total intensity, and is plotted on a square root scale to highlight the faint disk features. The linear Stokes parameters Q and U are plotted on a linear scale. The simulated dust grains do not induce circular polarization because we assume that the mechanisms that generate it, like multiple scattering or the presence of magnetic fields, are absent.}
    \label{fig:injected_stokes}
\end{figure*}

A debris disk model was constructed using the radiative transfer routine \texttt{MCFOST} \citep{pinte2006}. The model is intended to be an analogue of the HR 4796A debris disk system \citep[e.g.,][]{schneider1999hr4796a,perrin2015hr4796a}, with a host star ($T_{eff}\sim9250$ K) 73 pc away. The model was constructed to have a position angle of 28$\degr$, a vertical aspect ratio of 0.01, and a radius of peak dust density of 73.7 au. The radial density profile follows a two power-law structure, simulating a "sharp" inner profile ($\alpha_{\rm{in}} = 9.25$) cut-off to 0 at 60 au and a sharp outer profile ($\alpha_{\rm{out}} = -12.25$) cut-off at 20 au. We assume compact, spherical Mie-scattering grains \citep{mie1908} with 100\% astrosilicate composition, a minimum grain size of 1.8 $\mu$m, and a distribution of grain sizes following a single power-law function per \cite{dohnanyi1969collisional}. We assume that these dust grains don't generate circular polarization from physical processes like multiple scattering. We assume an observing wavelength of 860 nm ($I$-band). \texttt{MCFOST} generates I, Q, and U images of a disk model. To assess the effects of polarization aberrations on our final polarized intensity disk images, we simulate the model with the previously mentioned properties at an inclination of $i=75\degr$. Relative to the size of our detector, we generated smaller disk models to observe any significant effects from closer to the IWA of the coronagraph. In the highly-inclined disk case, the front-side of the disk (side closest to the observer) lies very near the IWA. The injected Stokes image of the disk is shown in Figure \ref{fig:injected_stokes}.

We construct the image of the disk subject to the GSMT polarization aberrations by transforming it via $\mathbf{M}_{coro}$ produced by the coronagraphic model used in Section \ref{sec:coronagraphy}. However, to account for the shift-
variant behavior near the inner working angle of the coronagraph we must simulate a $\mathbf{M}_{coro}$ for each point in the field of view that we are concerned with modeling and perform an integral transform on the injected Stokes image ($\mathbf{S}_{model}$) with $\mathbf{M}_{coro}$. We generate a library of $\mathbf{M}_{coro}$ evaluated at a number of points such that the injected Stokes image of the disk is Nyquist-sampled. The resulting field-dependent integral transform can be represented as sum of matrix-vector multiplications as shown in Equation \ref{eq:mvp},

\begin{equation}
    \mathbf{S}_{conv} (x, y) = \sum_{j = 0}^{N} \sum_{k = 0}^{N} \mathbf{M}_{coro, j, k} (x, y; \theta_{j, k}) \mathbf{S}_{model} (x, y),
    \label{eq:mvp}
\end{equation}

where $\mathbf{S}_{conv}$ is the simulated Stokes image with polarization aberration, $j, k$ are the indices of the grid of field points simulated along the $x, y$ axes, and $\mathbf{M}_{coro, j, k}$ is the MPRM evaluated at the $j-th$ and $k-th$ field point. We simulate the $\mathbf{S}_{conv}$ for each of the GSMTs using this disk model, and construct the normalized Stokes vector $\mathbf{s}_{conv}$ for each of them by dividing by the $I$ parameter as shown in Equation \ref{eq:normalized_stokes},

\begin{equation}
    \mathbf{s}_{conv} = \frac{1}{I_{conv} (x,y)} \mathbf{S}_{conv} = [1, q_{conv}(x, y), u_{conv}(x, y), v_{conv}(x, y)].
    \label{eq:normalized_stokes}
\end{equation}

\begin{figure*}
    \centering
    \includegraphics[width=0.75\textwidth]{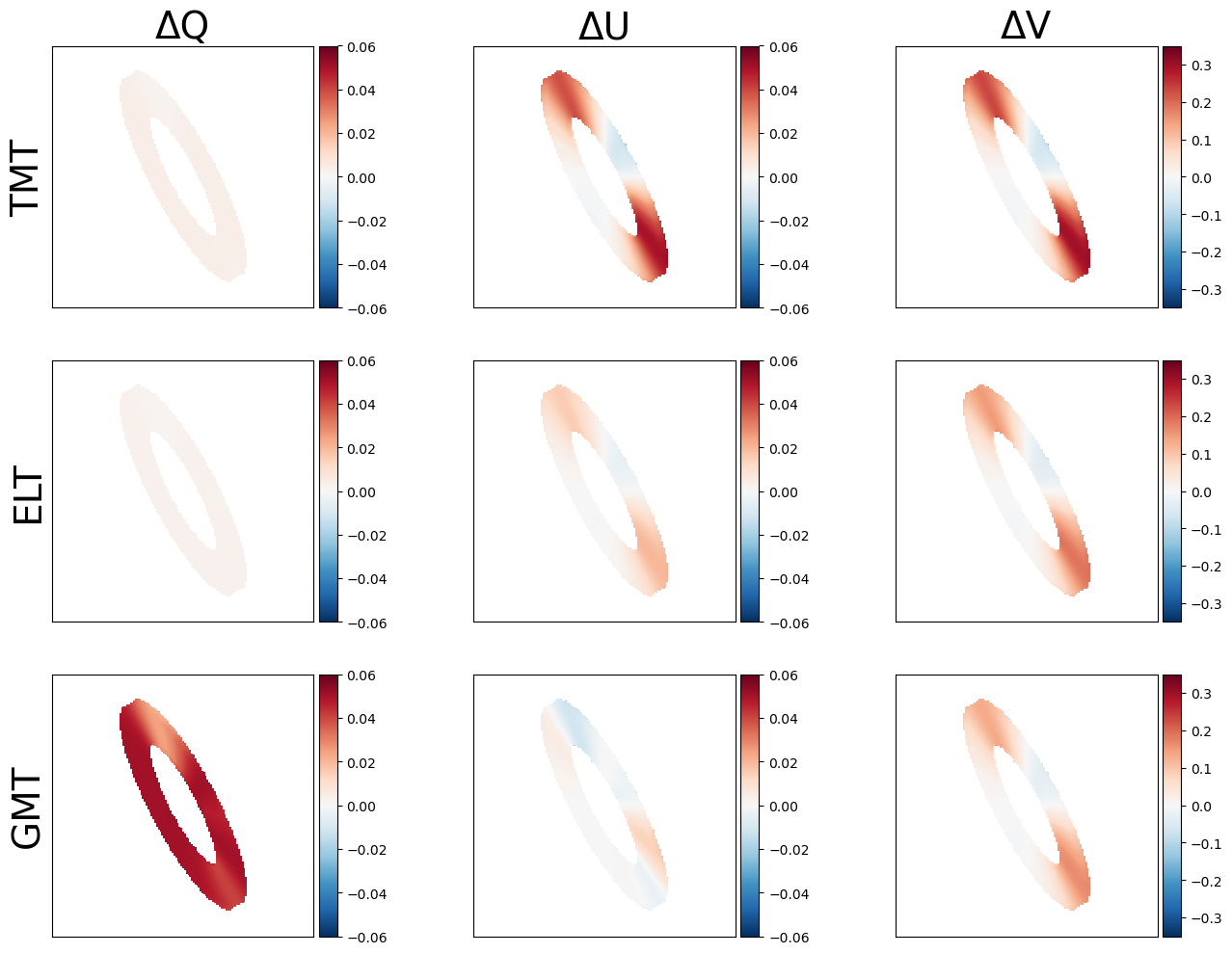}
    \caption{The difference in normalized Stokes parameters with respect to the debris disk model shown in Figure \ref{fig:injected_stokes}. We find that the GMT creates a large difference in the linear Stokes parameter $Q$, which is consistent with its large diattenuation in I-band that we reported on in \cite{GSMTsI_2023}. The largest change in the $U$ and $V$ Stokes parameters is from the TMT, which we know exhibits the greatest retardance in I-band. The $U$ difference images are symmetric, indicating a rotation of the detected polarization states observed from the disk. All three GSMTs suffer from a considerable degree of crosstalk, which is shown by the $V$ parameter which has zero input signal. This is readily apparent from the large $I \rightarrow V$ element of the Mueller PRM shown in Figure \ref{fig:tmt_prm}.}
    \label{fig:stokes_diff_images}
\end{figure*}

This normalizes the Stokes images to be expressed in terms of the degree of polarization for easy comparison that isn't a function of the intensity on a particular position on the disk. We then compute the difference of $\mathbf{s}_{conv}$ for each of the GSMTs with the normalized input Stokes image $\mathbf{s}_{model}$ to arrive at the difference in the Stokes parameters due to the polarization aberration model subject to segment-to-segment variations. These data are shown in Figure \ref{fig:stokes_diff_images}.

The lack of higher-order structure in these results shows that polarimetry is clearly dominated by the instrumental polarization and crosstalk of the GSMTs, rather than the segment-to-segment variation. The instrumental polarization of the GMT is highest, showing a strong $+Q$ polarization of the detected disk. This is consistent with the result in \cite{GSMTsI_2023} that the GMT is the greatest diattenuator. The linear-to-circular crosstalk is also strongest in the TMT, which we observed in \cite{GSMTsI_2023} to be the greatest retarder. The ELT experiences the least amount of both instrumental polarization and crosstalk, but has a similar overall structure to the TMT.  The maximum crosstalk experienced by the TMT is at roughly $30\%$ of the degree of polarization, which can substantially limit the sensitivity and accuracy of astronomical polarimeters by rotating the true polarization into the null space of the polarimeter, which typically senses linear light. The increased degree of $Q$ polarization is approximately constant for the GMT, and could in principle be calibrated out. The remaining symmetric errors that are prominent in the $\Delta U$ column are on the order of $\pm 4\%$, and may be indicative of retardance in the instrument that could also be calibrated out. Methods for doing so are covered exhaustively in the literature, which we report earlier in this section. However, the critical takeaway from this analysis is that the polarization aberrations from segment-to-segment variations are not a substantive limiter to polarimetry.

Note that the analysis conducted in this section does not consider the limitations that may be imposed by a bright host star. For example, faint disks will suffer from the residuals in polarized contrast introduced by the spatially-varying polarization near the inner working angle, like that shown in Figure \ref{fig:stokes_unpolarized_star}. Another interesting phenomenon to explore is how the polarized structure of the MPRM couples into scalar wavefront error, which has been observed on GPI in \cite{millar2022polarization}, and theoretically explored for the GSMTs in \cite{Ashcraft_2024_polabspeckle}. An accurate study of these effects is dependent on both the high-contrast imaging system, and target being observed, and will consequently be explored in future work.

%-------------------------------------- Two column figure (place early!)
   
% \subsection{Convolution with the Point Spread Functions}
% \subsection{PDI}
\section{Summary and Conclusion}
\label{sec:Summary}
A precise understanding of how polarization aberrations impact the GSMTs is critical for ground-based astronomy at shorter wavelengths. In this study, we present the next step required to bring models of polarization aberrations to higher fidelity. Below, we summarize the most critical findings from this study into polarization aberrations.

   \begin{enumerate}
      \item We have added another level of realism to our previously published polarization aberration models of the GSMTs that includes segment-to-segment variations.
      \item By assuming different spatial distributions of segment-to-segment coating variations on the primary mirror, we are able to statistically sample and parameterize the influence of nonuniform coatings on coronagraphic performance. For the middling case, we determined that the RMS contrast variation does not exceed $2 \times 10^{-8}$ contrast for the most sensitive coronagraph (order 2) and worst polarization aberrations (TMT) in I-band. This is orders of magnitude below the AO-limited residuals anticipated for the GSMTs, and will not be a substantive error term.
      \item Our analysis of the coronagraphic residuals is dependent on the assumptions made about the structure of the segment-to-segment variations. Little is known about the actual spatial structure of the coatings deposited on large segments, and their refractive indices and film thicknesses should be measured for a more accurate analysis. However, given the substantial variation in film thickness studied in this work, we do not expect segment-to-segment coating variations to be a limiter in high-contrast imaging.
      \item We assess the degree to which polarization aberrations from segment-to-segment variations will impact polarimetry of stars and debris disks. Ultimately we find that the coronagraphic Mueller point-response matrix exhibits polarized structure near the inner working angle of the coronagraphs. This means that observations of both unpolarized and polarized stars will have spatially-varying polarized structure near the inner working angle of a coronagraph. This effect does not appear to substantially alter the morphology of polarized off-axis structures like debris disks, whose detection errors are dominated by instrumental polarization and crosstalk.
      \item Our analysis of the influence on polarimetry could be expanded by incorporating these polarization aberration models into end-to-end models of coronagraphs, which will result in a polarized background that could limit polarimetric detection (using PDI) of faint objects like exoplanets.
      \item As part of this paper, we have built a GSMT polarization aberration forward modeling pipeline that is built using Poke's polarization ray tracing module and the HCIPy physical optics propagation package \citep{gsmts_sim}.
      \item While the goal of direct exoplanet detection and spectroscopy for the GSMTs may be constrained by the nominal polarization aberrations in the GSMTs, the influence of segment-to-segment coating thickness variations does not contribute a substantive error term to high contrast detection or polarimetry.
   \end{enumerate}

We plan to fully integrate our simulations described here into an end-to-end model of a high-contrast imaging instrument planned for the GSMTs. For example, the GMagAO-X instrument \citep{males2022conceptual} is an extreme-AO system in development that aims to achieve visible to near infrared high-contrast imaging on the GMT to search for planets in reflected light orbiting nearby stars. Similarly, the Planetary Camera and Spectrograph (PCS) is planned for the ELT, and the Planetary Systems Imager is planned for the TMT, both of which will cover similar wavelengths \citep{kasper2021pcs, Fitzgerald_2022}. 

Merging a realistic coronagraph model with our polarization aberration model, and integrating it with an end-to-end Fresnel analysis will be the next step toward understanding the precise influence of polarization aberrations from the GSMTs. The intermediate optics and atmospheric residuals will create a speckle field at the focal plane that, due to polarization aberrations, has some degree of polarization \citep{millar2022polarization, Ashcraft_2024_polabspeckle}. Studying how this effect can limit wavefront control algorithms is critical to optimizing the science yield of next-generation high-contrast imaging instruments.

\begin{acknowledgements}
    This work was supported by a NASA Space Technology Graduate Research Opportunity. The authors thank Saeko Hayashi, Ben Gallagher, and Warren Skidmore for helpful discussions about the planned TMT primary mirror segment coatings. S.Y.H acknowledges funding from the Heising Simons Foundations. J.N.A was supported by NASA through the NASA Hubble Fellowship grant \#HST-HF2-51547.001-A awarded by the Space Telescope Science Institute, which is operated by the Association of Universities for Research in Astronomy. This research made use of several open-source Python packages, including \emph{HCIPy} \citep{por2018hcipy}, \emph{prysm} \citep{Dube:22}, \emph{numpy} \citep{harris2020array}, \emph{matplotlib} \citep{Hunter:2007}, \emph{ipython} \citep{PER-GRA:2007}, and \emph{astropy} \footnote{http://www.astropy.org} \citep{astropy:2013, astropy:2018, astropy:2022}. 
\end{acknowledgements}


\bibliographystyle{aa} % style aa.bst
\bibliography{ref,ref_gsmts1}

\begin{appendix}

\section{RMS Contrast Variations}
\begin{figure*}
    \includegraphics[width=\textwidth]{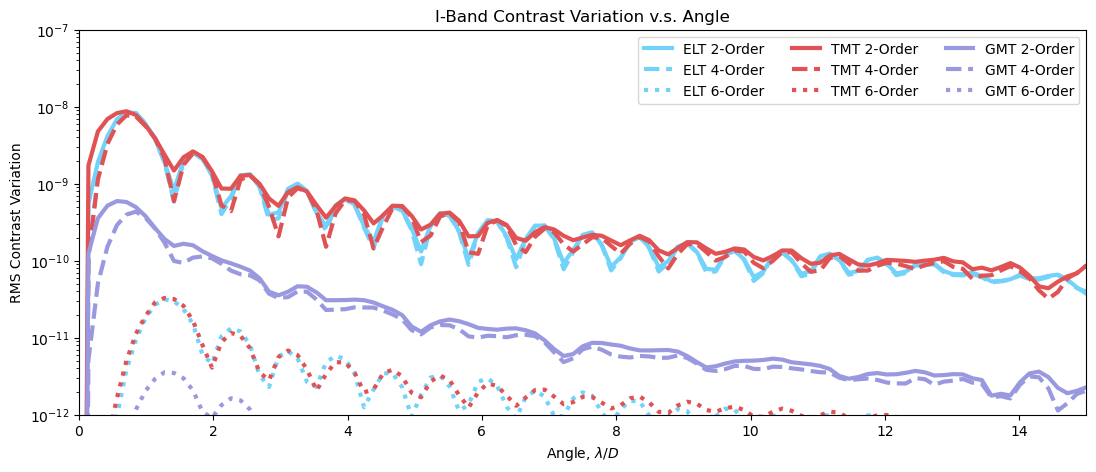}
    \caption{Plot of the RMS contrast variation as a function of angle for Case 1 in I-band. The lines represent the azimuthal average of the standard deviation of the simulated coronagraphic residuals. These trends tend to mimic the behavior in Figure \ref{fig:psf_residuals}, but be a factor of 2-5 lower in RMS contrast.}
    \label{fig:contrast_var_case1}
\end{figure*}

\begin{figure*}
    \includegraphics[width=\textwidth]{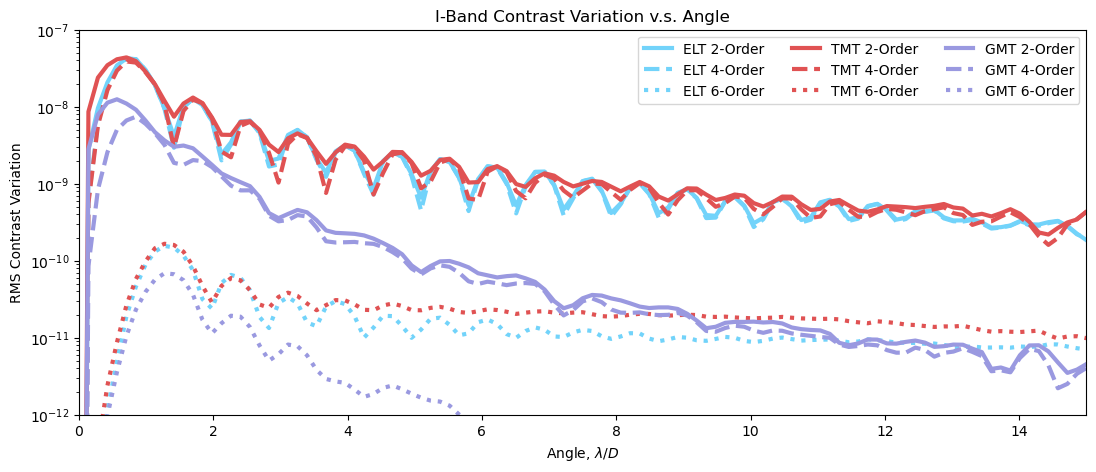}
    \caption{Plot of the RMS contrast variation as a function of angle for Case 3 in I-band. The lines represent the azimuthal average of the standard deviation of the simulated coronagraphic residuals. These trends tend to mimic the behavior in Figure \ref{fig:psf_residuals}, but be a factor of 2-5 lower in RMS contrast.}
    \label{fig:contrast_var_case3}
\end{figure*}

The RMS contrast variations for Case 1 (Figure \ref{fig:contrast_var_case1}) and Case 3 (Figure \ref{fig:contrast_var_case3}). As anticipated, Case 1 outperforms Case 2 (Figure \ref{fig:psf_residuals}), and Case 3 underperforms Case 2. Even for the worst-case contrast variation, the RMS Contrast does not exceed $10^{-7}$ contrast.

\label{sec:appendix_a}

\section{Mueller Point-Response Matrices for the GMT and ELT}
\label{sec:appendix_b}

\begin{figure*}
    \centering
    \includegraphics[width=\textwidth]{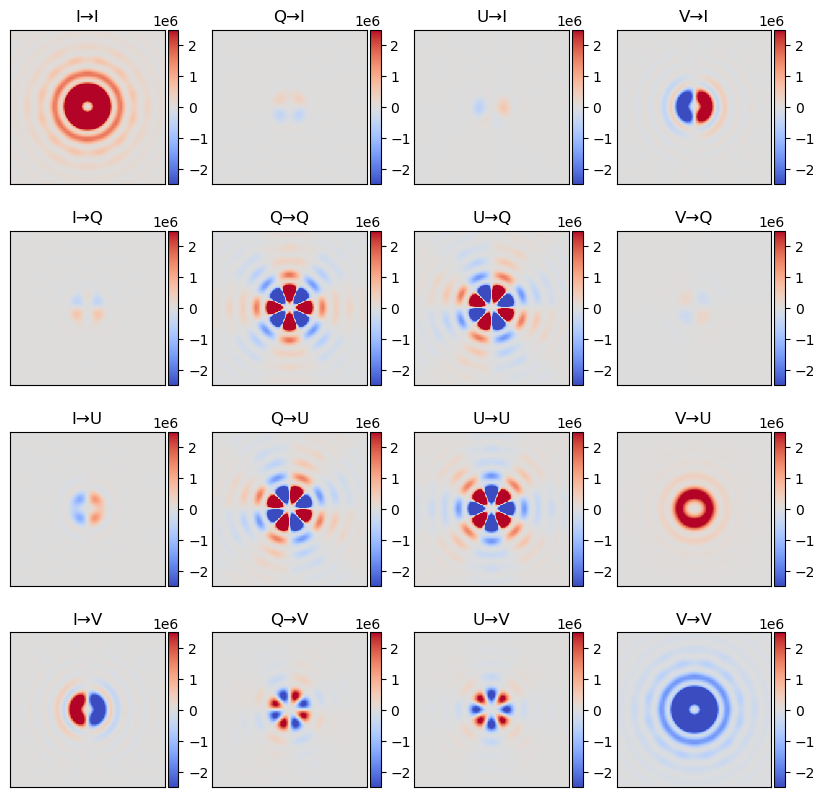}
    \caption{Figure showing the Mueller PRM $\mathbf{M_{coro}}$ for the ELT in I-band to show how polarization shapes the PSF. These data are normalized to the peak of the $I\rightarrow I$ component. The data here is plotted with a colorbar such that bright signals are saturated to enhance the structure in the first lobe of the PRM elements.}
    \label{fig:mpsm_elt}
\end{figure*}

\begin{figure*}
    \centering
    \includegraphics[width=\textwidth]{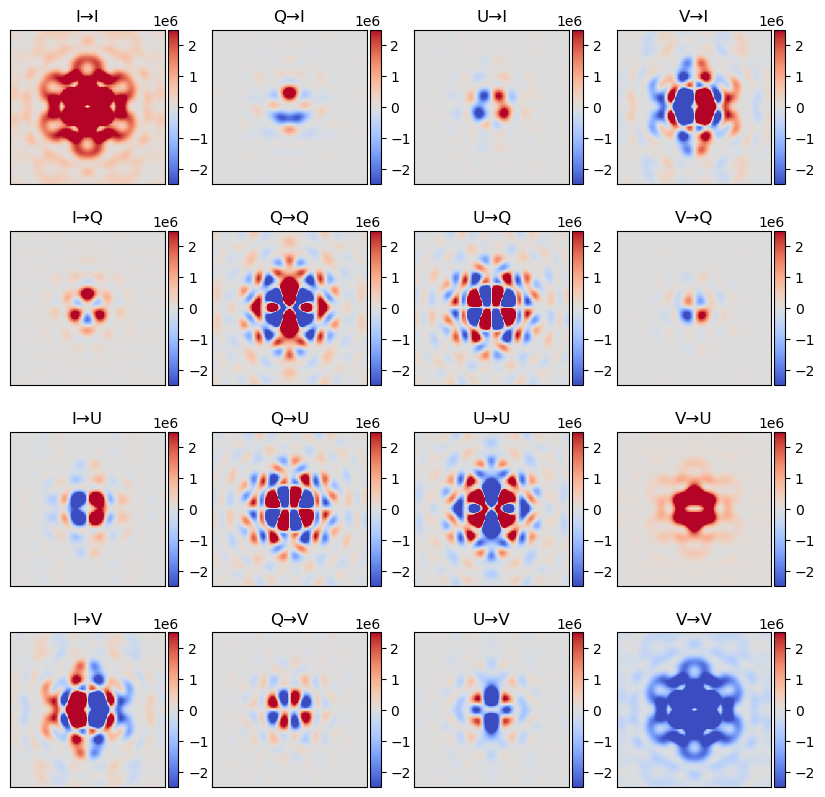}
    \caption{Figure showing the Mueller PRM $\mathbf{M_{coro}}$ for the GMT in I-band to show how polarization shapes the PSF. These data are normalized to the peak of the $I\rightarrow I$ component. The data here is plotted with a colorbar such that bright signals are saturated to enhance the structure in the first lobe of the PRM elements.}
    \label{fig:mpsm_gmt}
\end{figure*}

Figures illustrating the Mueller point-response matrices for the ELT (Figure \ref{fig:mpsm_elt}) and the GMT (Figure \ref{fig:mpsm_gmt}) to illustrate the polarized structure that can arise near the inner working angle of the perfect coronagraph models in the presence of polarization aberration.

\end{appendix}
\end{document}